\documentclass[aps,prl,twocolumn,showpacs,superscriptaddress,groupedaddress]{revtex4}

\usepackage{graphicx}
\usepackage{subfigure}
\usepackage{dcolumn}
\usepackage{bm}
\usepackage{amssymb}

\begin{document}

\title{Influence of small scale magnetic energy and helicity on the growth of large scale magnetic field}

\author{Kiwan Park$^{1}$\thanks{E-mail: pkiwan@gmail.com}}
 \altaffiliation{$^{1}$Ulsan National Institute of Science and Technology, South Korea}

\date{\today}

\begin{abstract}
The influence of initially given small scale magnetic energy($E_M(0)$) and helicity($H_M(0)$) on the magnetohydrodynamics(MHD) dynamo was investigated. Equations for $E_M$(t), $H_M$(t), and electromotive force($\langle {\bf v}\times {\bf b}\rangle$, $EMF$) were derived and solved. The solutions indicate small scale magnetic field(${\bf b}_i$) caused by $E_M$(0) modifies $EMF$ and generates additional terms of which effect depends on magnetic diffusivity $\eta$, position of initial conditions($IC$s) $k_f$, and time ($\sim e^{-\eta k_f^2 t}$). ${\bf b}_i$ increases the inverse cascade of energy resulting in the enhanced growth of large scale magnetic field($\overline{{\bf B}}$). Simulation data show that $E_M$(0) in small scale boosts the growth rate, which also proportionally depends on $H_M(0)$. If $E_M$(0) is the same, positive $H_M(0)$ is more effective for MHD dynamo than negative $H_M(0)$ is. It was discussed why large scale magnetic helicity should have the opposite sign of the injected kinetic helicity.
\end{abstract}

\maketitle

Most of the astrophysical objects commonly include the interactions between magnetic fields and conducting fluids. To figure out the celestial phenomena, understanding the evolutions of magnetic fields such as generation, amplification(dynamo), and annihilation(reconnection) are the most rudimentary; but, the mechanism is not well understood yet. Helical kinetic and magnetic fields which exist in the system without reflection symmetry are thought to play a key role in the evolution of magnetic field. The physical meaning and role of helical kinetic turbulent motion (kinetic helicity, $\langle {\bf u} \cdot {\bf \omega}\rangle$, ${\bf \omega}=\nabla \times{\bf u}$) are rather clear. It
transfers (kinetic) energy to magnetic eddy and cascades the (magnetic) energy to larger scale magnetic eddies. In contrast, the role of helical magnetic field i.e., magnetic helicity ($H_M\equiv\langle {\bf A} \cdot {\bf B}\rangle$, ${\bf B}=\nabla \times{\bf A}$) is partially known. $H_M$ is the topological measure of twist and linkage of magnetic field lines ($2\Phi_1\Phi_2$, $\Phi=\int_A {\bf B}\cdot d{\bf S}$, \cite{1980opp..bookR....K}, \cite{1978mfge.book.....M}), and is called the force free field(${\bf J} \times{\bf B}=0$, $\nabla \times{\bf B}=\lambda{\bf B}$) in the state of minimum energy equilibrium(\cite{2008matu.book.....B}). It has been known that $|H_M|$ in the large scale boosts the generation of ${\overline{B}}$ field(\cite{2013MNRAS.434.2020P}), but small scale $|H_M|$, more exactly $k^2|H_M|$, quenches the growth of ${\overline{B}}$ field. $H_M$ is part of $E_M$ and its magnitude cannot surpass $E_M/2k$(\cite{1975JFM....68..769F}). This realizability condition implies $E_M$ can have arbitrary magnetic helicity in the condition.\\

\noindent Another intuitional meaning of $H_M$ can be grasped from the distribution of $\langle B_i(k)B_j(-k)\rangle$(\cite{1987tfsn.book.....L}, \cite{2013MNRAS.434.2020P}, \cite{2011Springer.book.....Y}).
\begin{eqnarray}
\langle B_i({\bf k})B_j({\bf k'})\rangle = P_{ij}(k)\frac{E_M(k)}{4\pi k^2}+\frac{i\epsilon_{ijl}k_l}{8\pi k^2}H_M(k).\\
(P_{ij}(k)=\delta_{ij}-k_ik_j/k^2)\nonumber
\label{BiBj correlation}
\end{eqnarray}
This relation implies $H_M$ is basically related to the correlation between different components in case of a isotropic system without reflection symmetry. Like $E_M$ $H_M$ is a conserved quantity in an ideal MHD system, but it has a changeable sign, positive or negative. The unknown properties of $H_M$ within the degenerate $E_M$ leave room for investigating its influence on the profile of ${\bf B}$ fields and magnetic structure. Pouquet et al.(\cite{1976JFM....77..321P}) derived the equations of $E_M$, $H_M$ with kinetic equations and showed the inverse cascade of magnetic energy using EDQNM. But they chiefly concentrated on the derivation of $E_M$, $H_M$, and other variables paying less attention to their individual roles in MHD dynamo. Contrastively Ref.\cite{2013MNRAS.434.2020P} and \cite{2013PhRvE..87e3110P} showed the effect of initial $E_M$ or $H_M$ on MHD dynamo. However, the aims focused on the effect of $E_M$(0) rather than $H_M$(0). So we need another detailed analytic and experimental work that show the influence of $H_M$ coupled with $E_M$ more explicitly. For this, the equations of $E_M$ and $H_M$ were derived, and the solutions were found. Since $E_M$(0) and $H_M$(0) were given to the small scale magnetic eddy, the representation of modified $EMF$ was derived to explain simulation results.\\

For the simulation, magnetic field ${\bf b}_i$ with a fractional helicity($fhm$) drove a system as a precursor (one simulation step, t$<$0.005) to generate $E_M$(0) and $H_M$(0) at $k_f$=5. And then fully helical velocity field($fhk$=1.0) was injected to the kinetic eddy at $k_f$=5 (helical kinetic forcing $HKF$) as a main simulation. All simulations were done with high order finite difference Pencil Code(\cite{2001ApJ...550..824B}) and the message passing interface(MPI) in a periodic box of spatial volume $(2 \pi)^3$ with mesh size $256^3$. The basic equations solved in the code are,
\begin{eqnarray}
\frac{D \rho}{Dt}&=&-\rho \nabla \cdot {\bf u}\\
\frac{D {\bf u}}{Dt}&=&-c_s^2\nabla \mathrm{ln}\, \rho + \frac{{\bf J}\times {\bf B}}{\rho}+\nu\big(\nabla^2 {\bf u}+\frac{1}{3}\nabla \nabla \cdot {\bf u}\big)+{\bf f}\nonumber\\
\\
\frac{\partial {\bf A}}{\partial t}&=&{\bf u}\times {\bf B} -\eta\,\nabla\times{\bf B}.
\label{MHD equations in the code}
\end{eqnarray}
$\rho$: density; $\bf u$: velocity; $\bf B$: magnetic field; $\bf A$: vector potential; ${\bf J}$: current density;  $D/Dt(=\partial / \partial t + {\bf u} \cdot \nabla$): advective derivative; $\eta$: magnetic diffusivity(=$c^2/4\pi \sigma$, $\sigma$: conductivity); $\nu$: kinematic viscosity(=$\mu/\rho$, $\mu$: viscosity, $\rho$: density); $c_s$: sound speed. The unit used in the code is `$cgs$'. Velocity is expressed in units of $c_s$, and magnetic fields in units of $(\rho_0\,\mu_0)^{1/2}c_s$($B=\sqrt{\rho_0\,\mu_0}v$). $\mu_0$ is magnetic permeability and $\rho_0$ is the initial density. Note that $\rho_0\sim \rho$ in the weakly compressible simulations. These constants $c_s$, $\mu_0$, and $\rho_0$ are set to be `1'. In the simulations $\eta$ and $\nu$ are 0.006. To force the magnetic eddy($HMF$), forcing function `${\bf f}(x,t)$' is placed at Eq.(4) first; and then `${\bf f}$' is placed at Eq.(3) to drive the momentum equation($HKF$). ${\bf f}(x,t)$ is represented by $N\,{\bf f}_0(t)\, exp\,[i\,{\bf k}_f(t)\cdot {\bf x}+i\phi(t)]$($N$: normalization factor, ${\bf f}_0$: forcing magnitude, ${\bf k}_f(t)$: forcing wave number). The amplitude of magnetic forcing function($f_0$) was $0.01$ with various magnetic helicity ratios modifying $fhm$ during $HMF$; and $f_0$ of $HKF$ was $0.07$. $E_{kin}$ is not influenced by the short preliminary magnetic forcing so that all simulation sets have the same initial $E_{kin}$. As a result, $H_M$(0)/2 at $k$=5 are $\pm3.38\times10^{-6}$($fhm$=$\pm1$), $\pm2.34\times10^{6}$($fhm$=$\pm0.4$), $\pm1.31\times10^{-6}$($fhm$=$\pm0.2$), $-1.35\times10^{-10}$($fhm$=0). $E_M$(0) and $H_M$(0)/2 of reference simulation are $5.36\times 10^{-12}$ and $-4.68\times 10^{-14}$. However, $E_M(0)$ of each case is consistently the same: $1.82\times 10^{-5}$).\\

\begin{figure}
   {
     \includegraphics[width=8.0 cm]{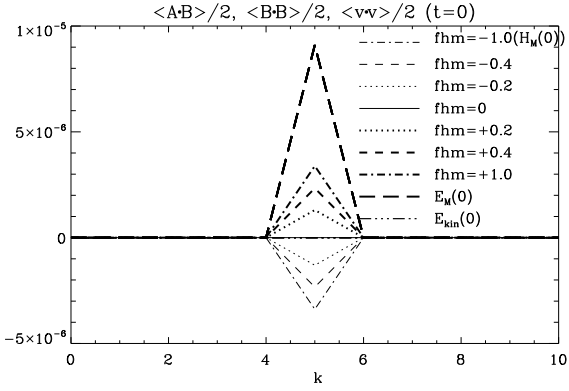}
     \label{1a}
}
\caption{$H_M(0)/2$ and $E_M(0)$ $k$=5. A thick long dashed line represents superposed $E_M$(0)s for all simulation sets. The horizontal line passing through `0' are $E_{kin}$(0)s for all simulations.}

\end{figure}

\begin{figure*}
\mbox{%
   \subfigure[]
   {
     \includegraphics[width=7.5cm]{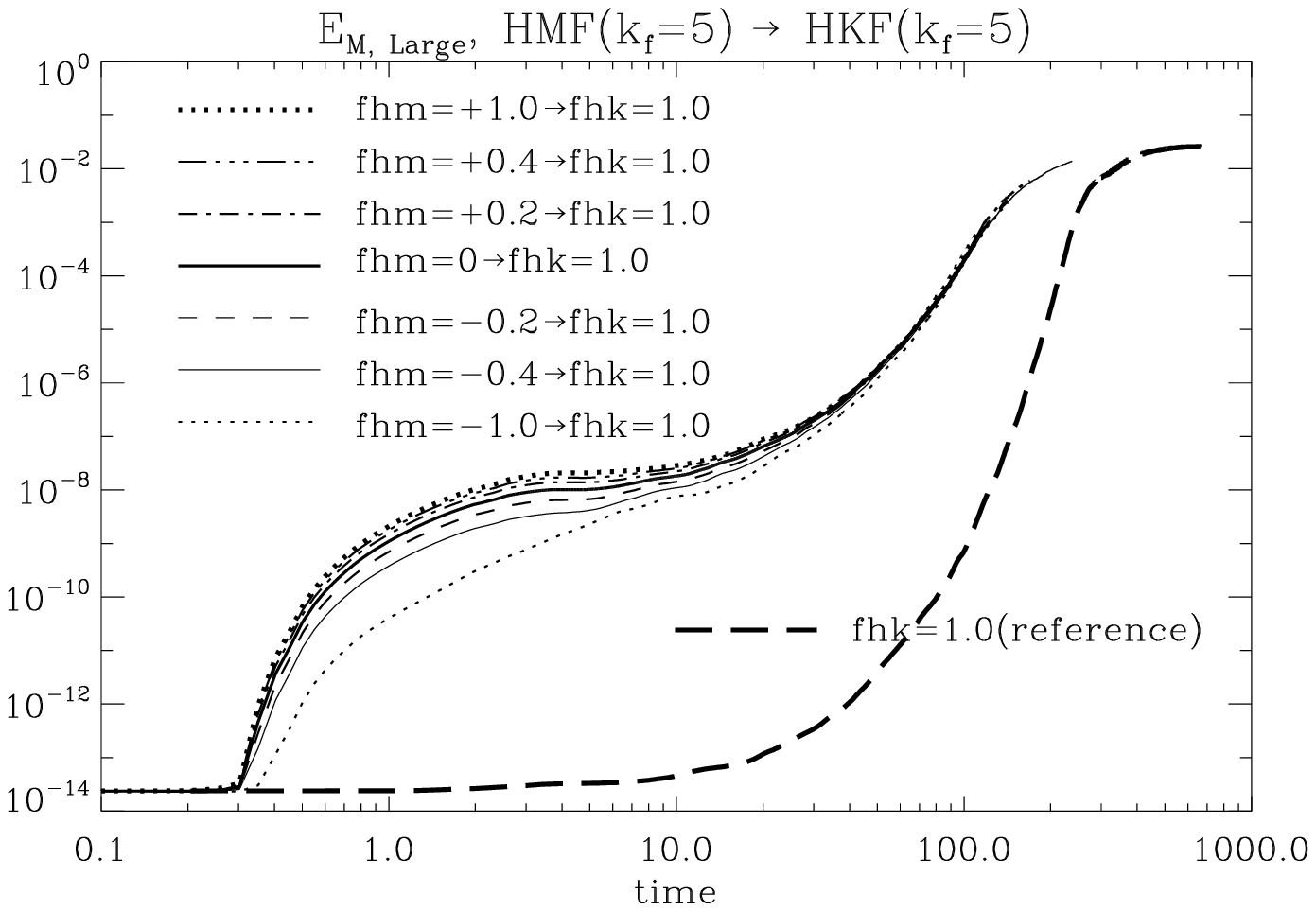}
     \label{2a}
               }\,
   \subfigure[]
   {
     \includegraphics[width=7.5cm]{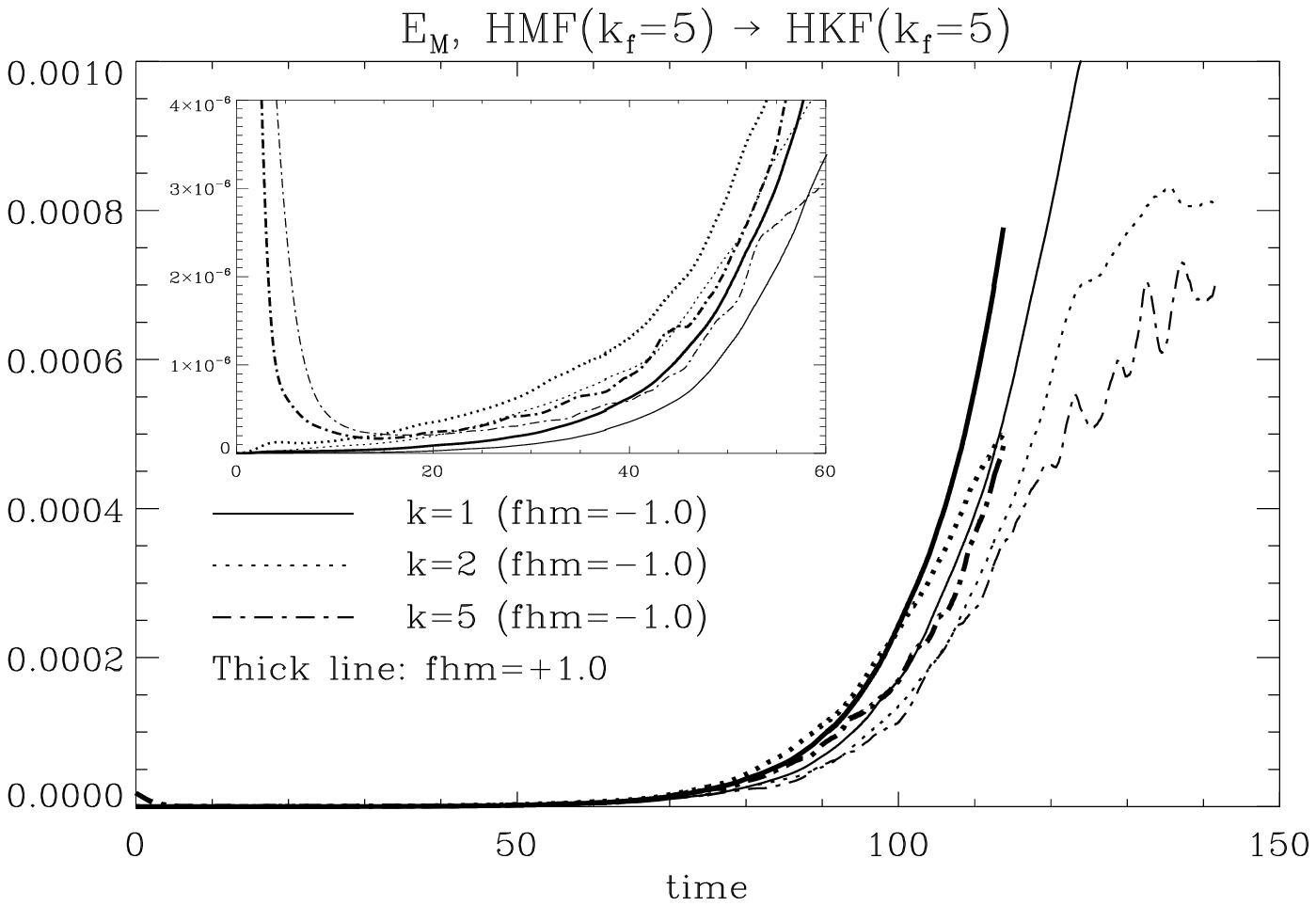}
     \label{2b}
               }
}

\mbox{
   \subfigure[]
   {
     \includegraphics[width=7.5cm]{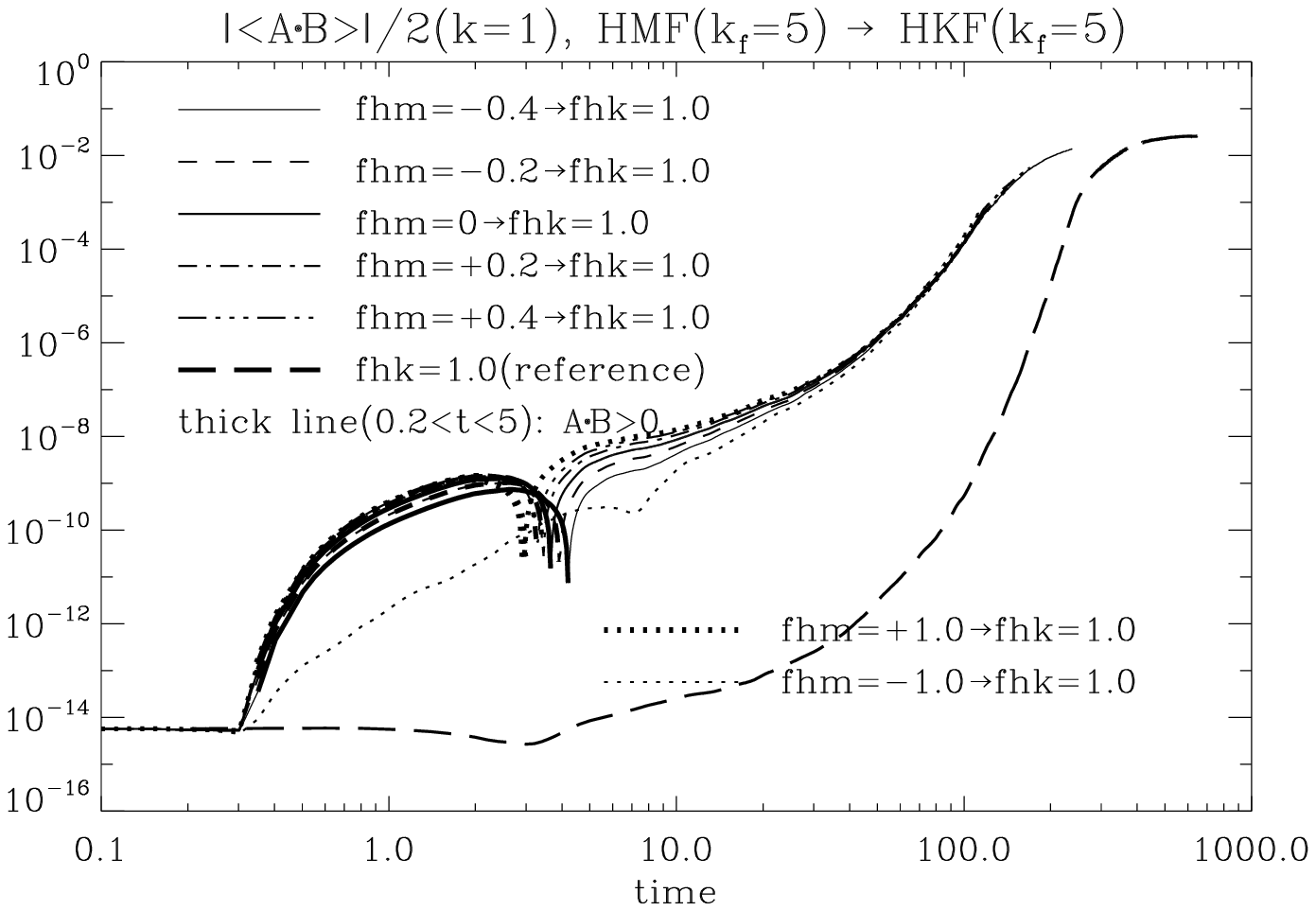}
     \label{2c}
               }\,
   \subfigure[]
   {
     \includegraphics[width=7.5cm]{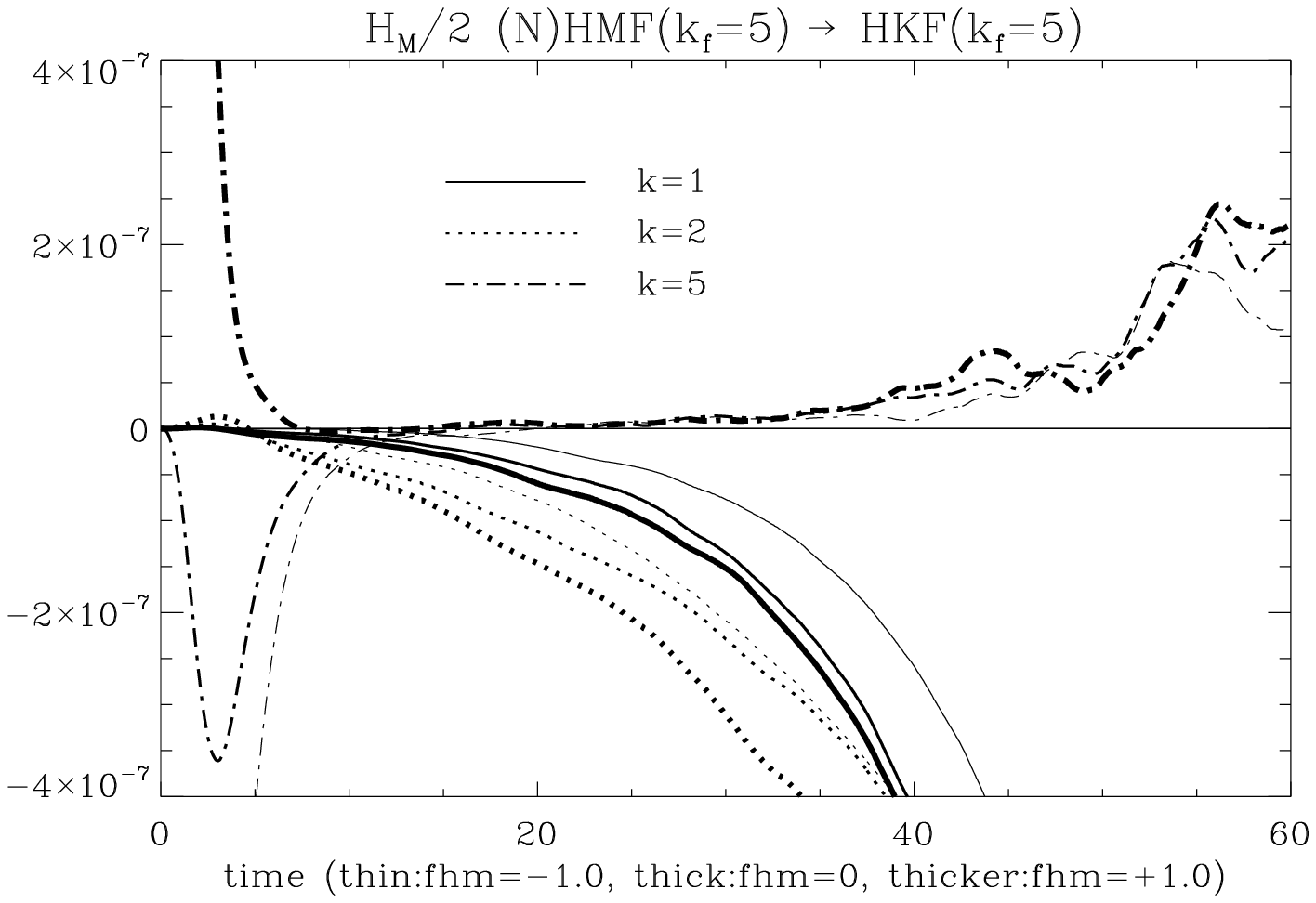}
     \label{2d}
               }

}

\caption{(a) The plots of $E_{M,L}$ with various $H_M$(0) at $k_f$=5. (b) $E_M$ at $k$=1, 2, and 5. Thin lines indicate $E_M$ with negative $H_M$(0)($fhm=-1$), thick lines are for $E_M$ with zero $H_M$(0)($fhm=0$), and thicker lines are for $E_M$ with positive $H_M$(0)($fhm=+1$). (c) $|H_{M,L}|$/2 with various $H_M$(0) at $k_f$=5. Most of $H_{M,L}$ is negative when the system is driven by the positive $\langle {\bf v}\cdot {\bf \omega}\rangle$, but $H_{M,L}$ is positive(thick lines) at $0.3<t<8$. (d) $H_M$ of $k$=1, 2, and 5.}

\end{figure*}

Instead of Eq.(2)-(4) we can use more simplified equations in an isotropic MHD system lacking reflection symmetry for the analytic approach. If we combine Faraday's law $\partial {\bf B}/\partial t=-c\nabla\times{\bf E}$ and Ohm's law ${\bf J}=\sigma({\bf E}+1/c{\bf U}\times {\bf B}$), we get the magnetic induction equation:
\begin{eqnarray}
\frac{\partial {\bf B}}{\partial t}=\nabla \times ({\bf U}\times {\bf B}) + \eta \nabla^2{\bf B}.
\label{magnetic induction 1}
\end{eqnarray}
If variables are split into the mean values and fluctuations like ${\bf U}={\bf \overline{U}}+{\bf u}\rightarrow {\bf u}$(${\bf \overline{U}}\equiv$0) and ${\bf B}={\bf \overline{B}}+{\bf b}$, the magnetic induction equation for ${\bf \overline{B}}$ field becomes(\cite{1980opp..bookR....K}),
\begin{eqnarray}
\frac{\partial {\bf \overline{B}}}{\partial t}&=&\nabla \times \langle {\bf u}\times {\bf b}\rangle + \eta \nabla^2{\bf \overline{B}}\\
&\sim&\nabla \times \alpha {\bf \overline{B}} + (\eta+\beta) \nabla^2{\bf \overline{B}}.
\label{magnetic induction 2}
\end{eqnarray}
(Here, electromotive force $EMF$ $\langle {\bf u}\times {\bf b}\rangle$ was replaced by $\alpha{\bf \overline{B}}-\beta \nabla \times {\bf \overline{B}}$ ($\alpha=1/3\int^t (\langle {\bf j}\cdot {\bf b}\rangle-\langle {\bf u}\cdot {\bf \omega}\rangle)d\tau,\,\, \beta=1/3\int ^t\langle u^2\rangle\,d\tau$ \cite{1978mfge.book.....M})\\
$E_M$(t) or $H_M$(t) can be found from the exact momentum and magnetic induction equation using EDQNM(\cite{1976JFM....77..321P}). But the same $H_M$(t) and $E_M$(t) can be derived from mean field method(\cite{2012MNRAS.419..913P}, \cite{2012MNRAS.423.2120P}). From Eq.(\ref{magnetic induction 2}), we get $\partial H_M / \partial t$(\cite{2002PhRvL..89z5007B}, \cite{1980opp..bookR....K}):
\begin{eqnarray}
\frac{\partial}{\partial t}\langle{\bf \overline{A}}\cdot {\bf \overline{B}}\rangle&=&
2\langle{\bf \overline{\xi}}\cdot {\bf \overline{B}}\rangle-2\eta \langle{\bf \overline{B}}\cdot\nabla \times {\bf \overline{B}}\rangle\nonumber\\
&=&2\alpha\langle {\bf \overline{B}}\cdot {\bf \overline{B}} \rangle-2(\beta+\eta)\langle {\bf \overline{B}}\cdot\nabla\times{\bf \overline{B}}\rangle
\label{Hm1}
\end{eqnarray}
In Fourier space
\begin{eqnarray}
\frac{\partial}{\partial t}H_{M,L}=4\alpha E_{M,L}-2k^2(\beta+\eta)H_{M,L}\,\,(k=1).
\label{Hm2}
\end{eqnarray}
Also $\partial_t E_{M,L}$ can be derived from Eq.(\ref{magnetic induction 2}).
\begin{eqnarray}
\frac{\partial }{\partial t}\frac{1}{2}\langle \overline{B}^2\rangle&=&\langle{\bf \overline{B}}\cdot\nabla\times {\bf\xi}\rangle-\frac{c}{\sigma}\langle{\bf \overline{B}}\cdot \nabla\times{\bf \overline{J}}\rangle\nonumber\\
&=&\langle\alpha {\bf \overline{B}}\cdot \nabla \times{\bf \overline{B}}\rangle-\langle\beta \nabla \times {\bf \overline{B}}\cdot \nabla \times{\bf \overline{B}}\rangle\nonumber\\
&&-\frac{c}{\sigma}\langle{\bf \overline{J}}\cdot \nabla\times{\bf \overline{B}}\rangle
\label{Em1}
\end{eqnarray}
In fourier space,
\begin{eqnarray}
\frac{\partial }{\partial t}E_{M,L}
&=&\alpha k^2\langle{\bf \overline{A}}\cdot {\bf \overline{B}}\rangle-k^2\big(\beta+\eta\big)\langle{\bf \overline{B}}^2\rangle\nonumber\\
&=&\alpha k^2H_{M,L}-2k^2\big(\beta+\eta\big)E_{M,L}\,(k=1)
\label{Em2}
\end{eqnarray}
$\partial E_M/\partial t$ and $ \partial H_M/\partial t$
have two normal mode solutions $\langle {\bf A}\cdot {\bf B}\rangle+\langle {\bf B}\cdot {\bf B} \rangle$ and $\langle {\bf A} \cdot {\bf B}\rangle-\langle {\bf B}\cdot {\bf B}\rangle$, from which two exact solutions can be derived(\cite{2013MNRAS.434.2020P}).
\begin{eqnarray}
2H_{M,L}(t)&=&(H_{M,L0}+2E_{M,L0})e^{2\int^t_0(\alpha-\beta-\eta)d\tau}\nonumber\\
&+&(H_{M,L0}-2E_{M,L0})e^{-2\int^t_0(\alpha+\beta+\eta)d\tau}
\label{EmHmSolution1}
\end{eqnarray}
\begin{eqnarray}
4E_{M,L}(t)&=&(H_{M,L0}+2E_{M,L0})e^{2\int^t_0(\alpha-\beta-\eta)d\tau}\nonumber\\
&-&(H_{M,L0}-2E_{M,L0})e^{-2\int^t_0(\alpha+\beta+\eta)d\tau}
\label{EmHmSolution2}
\end{eqnarray}
$E_{M,L}(t)$ and $H_{M,L}(t)$ proportionally depend on the initial values of large scale, which was shown in Ref.\cite{2013MNRAS.434.2020P}. Alfv$\acute{e}$n effect of large scale is so strong that the evolution of small scale ${\bf b}$ field is subordinate to the large scale magnetic field. The other factors affecting the growth of ${\bf \overline{B}}$ field are $\int_0^t (\alpha-\beta-\eta)\, d\tau$ and $\int_0^t (\alpha+\beta+\eta)\, d\tau$ originated from small scale velocity and magnetic field. However, the effect of initial small scale magnetic field ${\bf b}_i(0)$ distinctly appears only when $E_{M,L0}$ and $H_{M,L0}$ are small. Since $\alpha$ is negative and other terms are positive, the second terms in Eq.(\ref{EmHmSolution1}) and Eq.(\ref{EmHmSolution2}) dominantly decide the evolution of $E_{M,L}$ and $H_{M,L}$. Also negative `$H_{M,L0}-2E_{M,L0}$' indicates the evolving $E_M$ is positive but $H_{M,L}$ is negative.\\
We should note that $E_M$ in Eq.(\ref{EmHmSolution2}) is limited to the helical magnetic energy. 
If the system is not isotropic, a general case, the assumption that $\langle \bf{v}\times {\bf b}\rangle$ is the linear function of ${\bf \overline{B}}$ is hard to be used. $E_M$ is partially helical and partially nonhelical so quite a few considerable terms additively remain. Nonetheless, since the small scale field is globally inclined to be isotropic(locally anisotropic in MHD) and the system is driven by a fully helical kinetic energy, the assumption is valid and reasonable.\\

\begin{figure*}
\mbox{%
   \subfigure[]
   {
     \includegraphics[width=9.5cm]{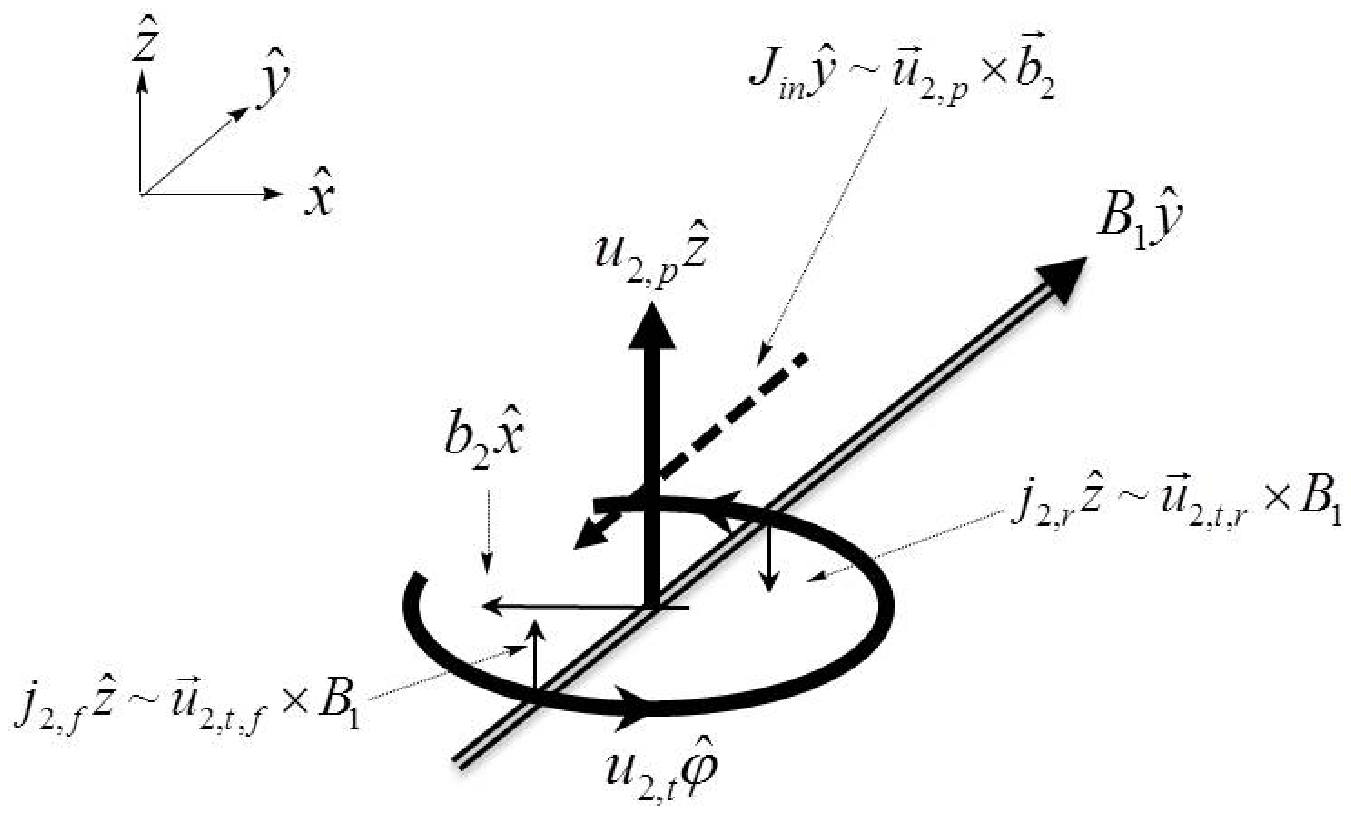}
     \label{4a}
               }\,
   \subfigure[]
   {
     \includegraphics[width=5.5cm]{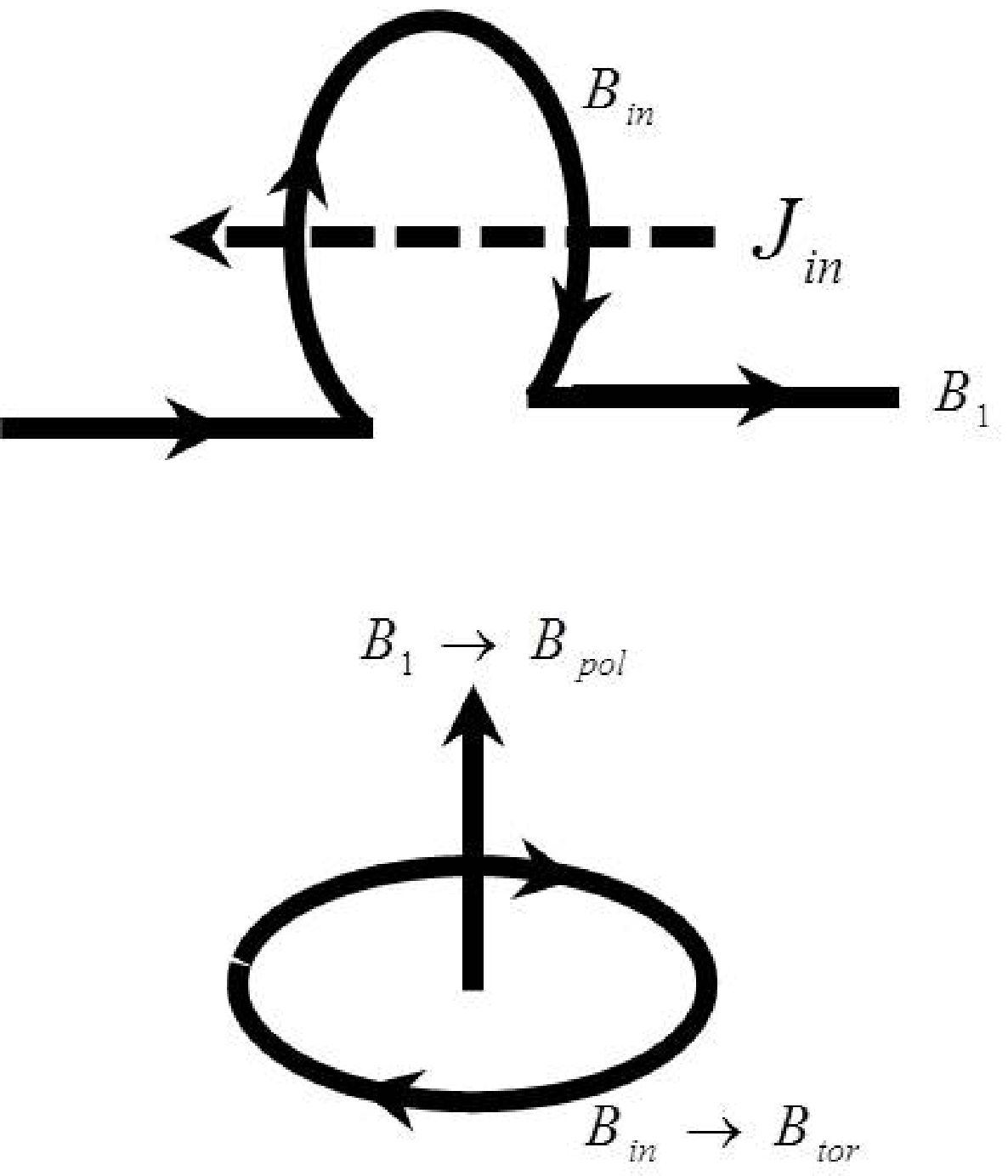}
     \label{4b}
               }

}
\caption{The generation of ${\bf J}_{in}$ is in consequence of the interaction between ${\bf u}$ and ${\bf B}_1$. ${\bf J}_{in}$ produces ${\bf B}_{in}$ from ${\bf B}_1$, both of which develop a left handed magnetic field structure.}

\end{figure*}

\noindent Our main interest if how the initial ${\bf b}_i(0)$ affects $EMF$ and the growth rate of large scale magnetic field eventually. We can guess ${\bf b}_i$(0) in small scale and this new field modifies $\alpha$ and $\beta$ in $EMF$. The magnetic induction equation due to $E_M$(0) is,
\begin{eqnarray}
\frac{\partial {\bf b}_i}{\partial t}
\approx\eta\nabla^2{\bf b}_i.
\label{MagIndSmall1}
\end{eqnarray}
In Fourier space,
\begin{eqnarray}
\frac{\partial {\bf b}_i}{\partial t}
\approx-\eta k_f^2{\bf b}_i\Rightarrow {\bf b}_i(t)={\bf b}_i(0)\,e^{-\eta k^2_ft}.
\label{MagIndSmall2}
\end{eqnarray}
While weak magnetic field does not conspicuously affect the velocity field ${\bf u}$, ${\bf u}$ determines the profile of ${\bf b}$ field. Ignoring dissipation term, the approximate small scale magnetic field ${\bf b}$ is
\begin{eqnarray}
\frac{\partial {\bf b}}{\partial t}
&=&\nabla\times({\bf u}\times\overline{\bf B})+\nabla\times({\bf u}\times{\bf b}_i).
\label{MagIndSmall3}
\end{eqnarray}
With the assumption of isotropy without mirror symmetry, $EMF$ $\xi$(=$\xi_1$+$\xi_2$, $\xi_1$ by ${\bf \overline{B}}$, $\xi_2$ by ${\bf b}_i$) can be represented by a simple form. Eq.(\ref{MagIndSmall3}) indicates ${\bf b}$ is a linear function of ${\bf B}$ and ${\bf b}_i$(kinetic dynamo). Hence, the components of $EMF$ will be only ${\bf B}$, $\nabla\times{\bf B}$, ${\bf b}_i$, $\nabla\times{\bf b}_i$ (we drop divergence of magnetic field). Thus, the basic structure of EMF is (\cite{2008matu.book.....B}, \cite{1980opp..bookR....K})
\begin{eqnarray}
\xi=\xi_1+\xi_2=\alpha_1 {\bf B}-\beta_1 \nabla \times{\bf B}+\alpha_2 {\bf b}_i-\beta_2 \nabla \times{\bf b}_i.
\label{modified EMF alpha beta}
\end{eqnarray}
Then $\xi_2$ is from,
\begin{eqnarray}
\xi_2&=&\int_{-\infty}^t \overline{u({\bf x}, t)\times\nabla\times
\big({\bf u}({\bf x}, t')\times{\bf b}_i({\bf x}, t')\big)}\,dt'
\label{EMF soulution2}
\end{eqnarray}
All the components of Eq.(\ref{EMF soulution2}) except the terms that fit Eq.(\ref{modified EMF alpha beta}) are to be dropped. Thus,
\begin{eqnarray}
\xi_{2x}&=&\bigg(u_y\frac{\partial u'_z}{\partial x}-u_z\frac{\partial u'_y}{\partial x}\bigg)b_{ix}-u_yu'_y\frac{\partial b_{iz}}{\partial y}
+u_zu'_z\frac{\partial b_{iy}}{\partial z}\nonumber\\
&\equiv& \alpha_{2x}b_{ix}-\beta_{2x}(\nabla\times b_i)_x
\label{simplified EMF x components}
\end{eqnarray}
$\xi_{2y}$ has the same structure but the variables rotate: $x$$\rightarrow$$y$, $y$$\rightarrow$$z$, $z$$\rightarrow$$x$. And for $\xi_{2z}$, $x$$\rightarrow$$z$, $y$$\rightarrow$$x$, $z$$\rightarrow$$y$.
Since small scale eddies are isotropic, the coefficients of $b_{ix}$, $b_{iy}$, $b_{iz}$ are the same.
\begin{eqnarray}
u_y\frac{\partial u'_z}{\partial x}-u_z\frac{\partial u'_y}{\partial x}=u_z\frac{\partial u'_x}{\partial y}-u_x\frac{\partial u'_z}{\partial y}=u_x\frac{\partial u'_y}{\partial z}-u_y\frac{\partial u'_x}{\partial z}\nonumber\\
\Rightarrow\frac{1}{3}\bigg(u_y\frac{\partial u'_z}{\partial x}-u_z\frac{\partial u'_y}{\partial x}+u_z\frac{\partial u'_x}{\partial y}-u_x\frac{\partial u'_z}{\partial y}+u_x\frac{\partial u'_y}{\partial z}-u_y\frac{\partial u'_x}{\partial z}\bigg)\nonumber\\
\label{simplified EMF alpha coefficient 1}
\end{eqnarray}
Then, $\alpha_2$ is
\begin{eqnarray}
\alpha_2=-\frac{1}{3}\int ^t_{-\infty}\overline{{\bf u}(x,t)\cdot \nabla \times {\bf u}(x,t')}\, dt'.
\label{alpha coefficient 2}
\end{eqnarray}
In the same way,
\begin{eqnarray}
u_xu'_x=u_yu'_y=u_zu'_z\Rightarrow \frac{1}{3}\big(u_xu'_x+u_yu'_y+u_zu'_z\big).
\label{simplified EMF beta coefficient 1}
\end{eqnarray}
\begin{eqnarray}
\beta_2=\frac{1}{3}\int ^t_{-\infty}\overline{{\bf u}(x,t)\cdot {\bf u}(x,t')}\, dt'
\label{alpha coefficient 2}
\end{eqnarray}

\noindent The coefficients $\alpha_1$ and $\beta_1$ related to $\overline{{\bf B}}$ are the same.\\

\noindent In contrast if the small scale magnetic field is so large that magnetic field meaningfully affects the velocity fields, we need to solve the momentum equation(${\bf u}$) first. We assume that dissipation effect is ignorably small and Lorentz force  is a dominant term in the momentum equation. Then,
\begin{eqnarray}
\frac{\partial {\bf U}}{\partial t}&\sim&{\bf J}\times {\bf B}=({\bf \overline{J}}+{\bf j}_i+{\bf j})\times ({\bf \overline{B}}+{\bf b}_i+{\bf b})
\label{alpha coefficient by magnetic field 1}
\end{eqnarray}
Small scale momentum equation is,
\begin{eqnarray}
\frac{\partial {\bf u}}{\partial t}&\sim&{\bf \overline{B}}\cdot\nabla{\bf b}+{\bf b}_i\cdot\nabla{\bf b}.
\label{alpha coefficient coefficient by magnetic field 3}
\end{eqnarray}
Here we assume the average of ${\bf b}_i$ is not zero and ${\bf b}_i$ does not change within the small scale eddy turnover time. $EMF$ is,
\begin{eqnarray}
{\bf u}\times {\bf b}&=&\int_{-\infty}^t{\bf \overline{B}}(x,t)\cdot\nabla{\bf b}(x,t')\,dt'\times {\bf b}(x,t)\nonumber\\
&&+\int_{-\infty}^t{\bf b}_i(x,t)\cdot\nabla{\bf b}(x,t')\, dt'\times {\bf b}(x,t)\nonumber\\
\label{alpha coefficient coefficient by magnetic field 4}
\end{eqnarray}
The integrand of $\xi_{M,1x}$ in the first term is,
\begin{eqnarray}
\xi_{M, 1x}\sim B_x\frac{\partial b'_y}{\partial x}b_z-B_x\frac{\partial b'_z}{\partial x}b_y.
\label{alpha coefficient coefficient by magnetic field x component}
\end{eqnarray}
$\xi_{M, 1y}$ and $\xi_{M, 1z}$ are the same with the rotation of variables mentioned above. And the assumption of isotropy makes the results simple.
\begin{eqnarray}
&&\frac{\partial b'_y}{\partial x}b_z-\frac{\partial b'_z}{\partial x}b_y=\frac{\partial b'_z}{\partial y}b_x-\frac{\partial b'_x}{\partial y}b_z=\frac{\partial b'_x}{\partial z}b_y-\frac{\partial b'_y}{\partial z}b_x\nonumber\\
&&\Rightarrow\frac{1}{3}\bigg(\frac{\partial b'_y}{\partial x}b_z-\frac{\partial b'_z}{\partial x}b_y+\frac{\partial b'_z}{\partial y}b_x-\frac{\partial b'_x}{\partial y}b_z+\frac{\partial b'_x}{\partial z}b_y-\frac{\partial b'_y}{\partial z}b_x\bigg)\nonumber\\
&&=\frac{1}{3}{\bf b}\cdot \nabla \times {\bf b'}
\label{alpha coefficient coefficient by magnetic field all components}
\end{eqnarray}
Thus, $\alpha_{M,1}$ related to ${\bf \overline{B}}$ field is
\begin{eqnarray}
\alpha_{M,1}=\frac{1}{3}\int ^t_{-\infty}\overline{{\bf b}(x,t)\cdot {\bf j}(x,t')}\, dt'\Rightarrow \alpha_{M,2}
\label{alpha coefficient due to magnetic field}
\end{eqnarray}
Finally, the complete $EMF$ is,
\begin{eqnarray}
\xi&=&\frac{1}{3}\big(\langle {\bf j}\cdot {\bf b} \rangle-\langle {\bf u}\cdot \omega \rangle \big)\tau{\bf \overline{B}}+\frac{1}{3}\langle u^2 \rangle \tau \nabla\times{\bf \overline{B}}\nonumber\\
&&+\frac{1}{3}\big(\langle {\bf j}\cdot {\bf b} \rangle-\langle {\bf u}\cdot \omega \rangle \big)\tau{\bf b}_i(0)e^{-\eta k_f^2t}\nonumber\\
&&+\frac{1}{3}\langle u^2 \rangle \tau \nabla\times{\bf b}_i(0)e^{-\eta k_f^2t}
\label{MagIndSmall4}
\end{eqnarray}
($\tau$ is substituted for the integration. Only magnitude is considered.)\\
${\bf b}_i$ makes the additional terms in $EMF$ causing $2\alpha \langle {\bf \overline{B}\cdot {\bf b}}_i\rangle\tau$ in Eq.(8) and $\alpha k_f^2\langle {\bf \overline{B}\cdot {\bf a}}_i\rangle\tau$ in Eq.(11). These new terms increase the growth rate of ${\bf \overline{B}}$. As Eq.(\ref{MagIndSmall2}) indicates, the influence depends on $\eta$, $k_f$, and time, which means the influence is gradually disappears. Those analytical results are supported by simulation.\\

Fig.1 shows the initial conditions $H_M$(0), $E_M$(0). This figure includes eight simulation sets of $fhm$=$\pm$1.0, $fhm$=$\pm$0.4, $fhm$=$\pm$0.2, and $fhm$=0. All simulations have the same $E_M$(0), and $E_{kin}$(0) after the preliminary $HMF$ is not influenced.\\

\noindent When the system with these initial conditions is driven by $HKF$, the growth of $E_{M,L}$ increases in proportion to $H_M(0)$(Fig.\ref{2a}). However, $E_M$(0) is in fact a more important factor in the growth rate. $E_M(0)$ and $H_M(0)$ of of the reference $HKF$ are $5.02\times10^{-12}$ and $-9.02\times10^{-14}$, and those for $fhm=-1$ are $1.82\times10^{-5}$ and $-6.77\times10^{-6}$. The plot shows that growth rate of the latter is even larger than that of the former.\\

\noindent Fig.\ref{2b} includes more detailed evolution of $E_M$ at $k$=1, 2, 5 for $fhm=0, \,\pm1.0$. The plot shows $E_M$ with the positive $H_M$(0) at $k$=5 decreases faster than that of negative $H_M$(0) when positive $\langle {\bf v} \cdot {\bf \omega}\rangle$(i.e., negative $H_M$ is produced) is injected to the system. This fast drop of $E_{M}$ at $k$=5 leads to the larger growth of $E_{M}$ at $k$=2, and this causes slow but larger growth rate of $E_{M,L}$.\\

\noindent Fig.\ref{2c} are to compare the growth rate of $|H_{M,L}|$ with various initial magnetic helicity. The growth rate is also proportional to $H_M$(0). Thick lines($0.3<t<8$) at Fig.\ref{2c} indicate $H_{M,L}$, therefor the cusp in the plot indicates the point(time) when the positive $H_M$ turns into a negative. This positive $H_M$ is thought to be caused by the tendency of conserving $H_{M, tot}$ against the injected negative $H_M$. However, if the ratio of negative $H_M$(0) to $E_M$(0) is large($fhm=-1$) enough or $E_M$ is not so large(reference $HKF$), $H_{M,L}$ does change a sign.\\

\noindent Fig.\ref{2d} shows the evolving profiles of $H_M$ at $k$=1, 2, and 5. $H_M$ at $k$=1, 2 grow to be negative, but $H_M$ of $k$=5 turns into a positive regardless of $H_M$(0). This is a sort of reactive interaction with the negative larger scale, i.e., conservation of $H_{M,tot}$. While $H_M$(0) at $k$=5 is positive, $H_M$(0) decreases faster than the negative $H_M$(0) does. And this fast decrease of $H_M$ boosts the growth of $|H_M|$ at $k$=1, 2. In contrast, for the negative $H_M$(0) at $k$=5, injected negative $H_M$ mitigates the decreasing speed of $|H_M|$ at $k$=5 and growing speed of $|H_M|$ at $k$=1, 2. In case $H_M$(0) at $k$=5 is 0, $H_M$ at $k$=5 first drops because the negative $H_M$ flows in. But as the magnitude of $\overline{B}$ field grows, the diffusion of positive $H_M$ from large scale, sort of counteraction on the basis of $H_{M, tot}$ conservation, makes $H_M$ at $k$=5 grow to be positive.\\

%
%

Up to now we have used the fact that the left handed magnetic helicity($\langle {\bf a}_2\cdot {\bf b}_2\rangle$$<0$) is generated when the system is driven by the right handed kinetic helicity($\langle {\bf u}_2\cdot {\bf \omega}_2\rangle>0$) without enough consideration. Mathematically the growth of larger scale magnetic field(${\bf B}_1$) or helicity(${\bf H}_1$) is described by a differential equation like Eq.(\ref{magnetic induction 2}) or Eq.(\ref{Hm2}). However, since the differential equation in itself cannot describe the change of sign of variables, more fundamental and physical approach is required.\\

\noindent In case of $\alpha$$\Omega$ dynamo, there was a trial to explain the handedness of twist and writhe in corona ejection using the concept of magnetic helicity conservation(\cite{2003ApJ...584L..99B}). However, even when there is no effect of differential rotation, $\alpha^2$ dynamo, the sign of generated $H_M$ is opposite to that of the injected $\langle {\bf u}\cdot {\bf \omega}\rangle$.\\

\noindent We assume the magnetic field $B_{1}\hat{y}$(Fig.\ref{4a}, \cite{1980opp..bookR....K}) interacts with right handed helical kinetic plasma motion. The velocity can be divided into toroidal component $u_{2,t}\hat{\phi}$ and poloidal component $u_{2,p}\hat{z}$. The interaction of this toroidal motion with ${\bf B}_{1}$ produces ${\bf j}_{2}$$\sim{\bf u}_{2,t}\times {\bf B_{1}}$. The induced current density ${\bf j}_{2,f}$ in the front is toward positive $\hat{z}$ direction, but the rear current density ${\bf j}_{2,r}$ is along with the negative $\hat{z}$. These two current densities become the sources of magnetic field $-{b}_{2}$$\hat{x}$(${\bf j}\sim \nabla \times {\bf b}$). Again this induced magnetic field interacts with the poloidal kinetic velocity $u_{2,p}\hat{z}$ and generates -$J_{in}\hat{y}$($\sim{\bf u}_{2,p}\times {\bf b_{2}}$). Finally this ${\bf J}_{in}$ produces ${\bf B}_{in}$, which forms a circle from the magnetic field ${\bf B}_1$(upper picture in Fig.\ref{4b}). If we go one step further from here, we see ${\bf B}_{in}$ can be considered as a new toroidal magnetic field ${\bf B}_{tor}$, and ${\bf B}_1$ as ${\bf B}_{pol}$. This new helical magnetic field structure has the left handed polarity, i.e., $\langle {\bf a}\cdot {\bf b}\rangle$$<$0 (lower plot of Fig.\ref{4b}). ${\bf B}_{tor}$ interacts with the positive $\langle {\bf u}_2\cdot {\bf \omega}_2\rangle$ and induces the current density ${\bf J}$ which is antiparallel to ${\bf B}_{tor}$. Then ${\bf B}_{pol}$ is reinforced by this ${\bf J}$. This is $\alpha^2$($B_{pol}\leftrightarrow B_{tor}$) dynamo with the external forcing source.\\

We have seen how $E_M$(0) and $H_M$(0) in small scale affect the growth of large scale $B$ field. 
The initially given magnetic energy produces additional terms that include ${\bf b}_i\sim e^{-\eta k^2_ft}$ in $EMF$. Roughly there appears to be a linear like relation between the growth rate of $E_{M,L}$ and $H_M$(0) or $E_M$(0). However, we need more study before making a conclusion because $E_M$ can involve arbitrary $H_M$($\leq 2E_M/k$) which is almost a conserved quantity with a changeable sign and magnitude. $E_M$(0) modifies $EMF$ term generating ${\bf b}_i$ and the related additional terms. Besides the three explicit factors($\eta$, $k_f$, $t$) affecting the growth rate, the interaction between ${\bf \overline{B}}$ or ${\bf \overline{A}}$ and ${\bf b}_i$ needs to be found out for more complete understanding the influence of $H_M$(0) and $E_M$(0). And deriving the modified $EMF$, we assumed the system was isotropic. But, it is rather difficult to use the assumption when there is a mean or large scale magnetic field in the system. At least we need to consider the local anisotropy due to the mean magnetic field although the system driven by isotropic random force is globally isotropic. We will leave this topic for the future work.\\

\noindent Kiwan Park acknowledges the grant and comments from Dr. Dongsu Ryu at UNIST.

\bibliography{bibdatabase}
\bibliographystyle{plain}

\end{document}